\documentclass[a4paper,12pt]{article}
\usepackage[pctex32]{graphics}
\usepackage{amssymb,amsmath}

\begin{document}
\topmargin 0pt
\oddsidemargin 0mm
\newcommand{\be}{\begin{equation}}
\newcommand{\ee}{\end{equation}}
\newcommand{\ba}{\begin{eqnarray}}
\newcommand{\ea}{\end{eqnarray}}
\newcommand{\fr}{\frac}
\renewcommand{\thefootnote}{\fnsymbol{footnote}}

\begin{titlepage}

\vspace{5mm}
\begin{center}
{\Large \bf Phase transitions for the Lifshitz black holes }

\vskip .6cm
 \centerline{\large
 Yun Soo Myung$^{a}$}

\vskip .6cm

{$^{1}$Institute of Basic Science and School of Computer Aided
Science,
\\Inje University, Gimhae 621-749, Korea \\}
\end{center}

\begin{center}

\underline{Abstract}
\end{center}
We study  possibility of  phase transitions between Lifshitz black
holes and other configurations by using  free energies explicitly. A
phase transition between Lifshitz soliton
 and Lifshitz black hole might not occur in three
dimensions. We find that  a phase transition between Lifshitz and
BTZ black holes unlikely occurs because they have different
asymptotes.  Similarly, we point out that any phase transition
between Lifshitz and black branes unlikely occurs in four dimensions
since they have different asymptotes. This is consistent with a
necessary condition for taking a phase transition in the
gravitational system,  which requires the same asymptote.

\vspace{5mm}


\vskip 0.8cm

\vspace{15pt} \baselineskip=18pt \noindent $^a$ysmyung@inje.ac.kr

\thispagestyle{empty}
\end{titlepage}

\newpage
\section{Introduction}
Recently, the Lifshitz-type black
holes~\cite{CFT-4,L-1,AL-3,L-2,L-4,L-3,L-5} have received
considerable attentions since these may provide a model of
generalizing AdS/CFT correspondence to non-relativistic condensed
matter physics as the AdS/CMT
correspondence~\cite{CFT-1,CFT-2,CFT-3}.  Although their asymptotic
spacetimes  are apparently  known to be Lifshitz, obtaining an
analytic solution seems to be a nontrivial task. The known solutions
include  a four-dimensional topological black hole which is
asymptotically Lifshitz with the dynamical exponent
$z=2$~\cite{Mann}.  Analytic black hole solution with $z=2$ that
asymptotes planar Lifshitz spacetimes was found  in the
Einstein-scalar-massive vector theory~\cite{bm} and  in the
Einstein-scalar-Maxwell theory~\cite{tay}. Another analytic solution
has  been recently found in the Lovelock gravity~\cite{MTr}. The
$z=3$ Lifshitz black hole~\cite{z3} was derived from the new massive
gravity (NMG) in three-dimensional spacetimes~\cite{bht}. Numerical
solutions to Lifshitz black hole and thermodynamic property of this
system were also explored in~\cite{BBP,AL-2}.

Their  thermodynamic studies  was limited because it was difficult
to compute their conserved quantities in asymptotic Lifshitz
exactly. Recently, there was a progress on computation of mass and
related thermodynamic quantities by using the ADT
method~\cite{DS-1,DS-2} and Euclidean action approach~\cite{GTT}.
Concerning the mass of Lifshitz black hole in three dimensions,
however,  there is a discrepancy between ${\cal
M}=\frac{7r_+^4}{8G_3 \ell^4}$ obtained from the ADT
method~\cite{DS-1} and ${\cal M}=\frac{r_+^4}{4G_3 \ell^4}$ from
other approaches~\cite{HT,MKP,GTT}. In this work, we use the latter
expression because it respects the first-law of thermodynamics and
the ADT mass is not reliable  to use for a thermodynamic study of
the Lifshitz black hole~\cite{DS-2}.

On the other hand,  the Schwarzschild black hole
  is  in an unstable equilibrium with the  heat
reservoir of the temperature $T$~\cite{GPY}. Its fate under small
fluctuations will be either to decay to  hot flat space by Hawking
radiation or to grow without limit by absorbing thermal radiations
in the infinite heat reservoir~\cite{York}. This means that an
isolated black hole is never in thermal equilibrium in
asymptotically  flat spacetimes. Thus,  one has to find  a way  of
achieving a stable black hole which is in an equilibrium with the
finite heat reservoir. A black hole could be rendered
thermodynamically stable by placing it in four-dimensional anti-de
Sitter (AdS$_4$) spacetimes because AdS$_4$ spacetimes plays the
role of a confining  box. An important point to understand is to
know how a stable black hole with positive specific heat could
emerge from thermal radiation through a phase transition. This was
known to be the Hawking-Page phase transition between thermal AdS
space and Schwarzschild-AdS black hole~\cite{HP,BCM}, which shows a
typical example of the {\it first-order phase transition} in the
gravitational system.  See Appendix for a detail description of how
the first-order phase transition occurs.

 Witten ~\cite{Witt} has extended this
four-dimensional transition to arbitrary dimension and provided a
natural explanation of a confinement/deconfinement transition on the
boundary field theory via the AdS/CFT correspondence.  On later, it
was proposed that a transition between black hole with scalar hair
(Martinez-Troncoso-Zanelli black hole~\cite{MTZ}) and a topological
black hole is possible to occur as a {\it second-order phase
transition}  in asymptotically AdS$_4$ spacetimes~\cite{KMPS}. We
have shown that the phase transition between these is second-order
when employing the temperature matching and using the difference of
free energies~\cite{Myungsh}.

Concerning  the Lifshitz black holes, it was firstly proposed that
the transition between Lifshitz and black branes may occur because
the anisotropic background of Lifshitz brane is favored at low
temperature, while the AdS background of black brane is favored in
high temperature~\cite{tay}. If it is possible to occur, it
corresponds to a phase transition between two different asymptotes
and inquiring its order is a curious question. Also,  it was
suggested  that the Lifshitz black hole found numerically in string
theory may appear from thermal Lifshitz via the Hawking-Page phase
transition at critical temperature~\cite{AF}.  Recently, three U(1)
fields extension of Einstein-scalar-Maxwell theory provided a
charged Lifshitz black hole and phase transitions between Lifshitz
black hole and thermal Lifshitz were discussed in~\cite{TV}. Very
recently, the Lifshitz soliton was proposed to be a ground state of
Lifshitz spacetimes in three dimensions~\cite{GTT}, implying that a
phase transition between
 Lifshitz soliton and Lifshitz black hole may occur,  as the
transition between thermal AdS$_3$ and BTZ black hole was possible
to occur~\cite{myungbtz}. However, {\it a necessary condition for a
phase transition between two configurations to take place is that
they should have the same asymptotic structure.}

In this work, in order to answer to the above questions partly,  we
wish to study possibility of phase transitions between Lifshitz
black holes and other configurations explicitly. Thermodynamics
study  is a key analysis for  all Lifshitz black holes.  If phase
transitions between Lifshitz black holes and other configurations
really occur, these would provide some information on  phase
transitions in condensed matter physics via the presumed AdS/CMT
correspondence.

The organization of our work is as follows. In section 2, we first
review a well-known phase transition between thermal AdS$_3$ and BTZ
black hole briefly. Then, we investigate a transition between
Lifshitz soliton with  $z=1/3$ and  Lifshitz black hole with $z=3$.
 In section 3, we study possibility of
transitions between Lifshitz brane with $z=2$ and black brane with
$z=1$.   It turns out that these all transitions are not  allowed
because two black holes have different asymptotes. Finally, we
discuss our results in section 4.

\section{Phase transitions in three dimensions}

The NMG action~\cite{bht} composed of the Einstein-Hilbert action
with a cosmological constant $\lambda$ and higher-order curvature
terms is given by
\begin{eqnarray}
\label{NMGAct}
 S^{(3)}_{NMG} &=&-\Big[ S^{(3)}_{EH}+S^{(3)}_{HC}\Big], \\
\label{NMGAct2} S^{(3)}_{EH} &=& \frac{1}{16\pi G_3} \int d^3x \sqrt{-g}~ ({\cal R}-2\lambda),\\
\label{NMGAct3} S^{(3)}_{HC} &=& -\frac{1}{16\pi G_3m^2} \int d^3x
            \sqrt{-g}~\left({\cal R}_{\mu\nu}{\cal R}^{\mu\nu}-\frac{3}{8}{\cal R}^2\right),
\end{eqnarray}
where $G_3$ is a three-dimensional Newton constant and $m^2$ a
parameter with mass dimension 2.  We  mention that to avoid negative
mass and entropy, it is necessary to take ``$-$" sign in the front
of $[ S^{(3)}_{EH}+S^{(3)}_{HC}]$.  The field equation is given by
\be {\cal R}_{\mu\nu}-\frac{1}{2}g_{\mu\nu}{\cal R}+\lambda
g_{\mu\nu}-\frac{1}{2m^2}K_{\mu\nu}=0,\ee where
\begin{eqnarray}
  K_{\mu\nu}&=&2\square {\cal R}_{\mu\nu}-\frac{1}{2}\nabla_\mu \nabla_\nu {\cal R}-\frac{1}{2}\square{\cal R}g_{\mu\nu}\nonumber\\
        &+&4{\cal R}_{\mu\nu\rho\sigma}{\cal R}^{\rho\sigma} -\frac{3}{2}{\cal R}{\cal R}_{\mu\nu}-{\cal R}_{\rho\sigma}{\cal R}^{\rho\sigma}g_{\mu\nu}
         +\frac{3}{8}{\cal R}^2g_{\mu\nu}.
\end{eqnarray}
In order to obtain  Lifshitz black hole solution with dynamical
exponent $z$, it is convenient to introduce dimensionless parameters
\be y=m^2~ \ell^2,~~w=\lambda~ \ell^2, \ee where $y$ and $w$ are
proposed to take \be y=-\frac{z^2-3z+1}{2},~~w=-\frac{z^2+z+1}{2}.
\ee In order to obtain the $z=1$  BTZ black hole~\cite{BTZ-1,BTZ-2},
one has $y=\frac{1}{2}$ and $w=-\frac{3}{2}$, while
 $y=-\frac{1}{2}$ and
$w=-\frac{13}{2}$ are chosen for getting the  $z=3$ Lifshitz black
hole.

Explicitly, we find the $z$-dependent black hole solution~\cite{z3}
as
 \be
\label{2dmetric}
  ds^2_{z}=g_{\mu\nu}dx^\mu dx^\nu=-\left(\frac{r^2}{\ell^2}\right)^z\left(1-\frac{M\ell^2}{r^2}\right)dt^2
   +\frac{dr^2}{\left(\frac{r^2}{\ell^2}-M\right)}+r^2d\phi^2,
\end{equation}
where $M$ is an integration constant related to the the  mass of
black hole.  The horizon radius $r_+$ is determined by the relation
of $g^{rr}=0$ and $\ell$ denotes the curvature radius of Lifshitz
(AdS$_3$) spacetimes. This line element is invariant under the
anisotropic scaling of \be t\to \tilde{\lambda}^zt,~~\phi \to
\tilde{\lambda} \phi, r\to \frac{r}{\tilde{\lambda}},~~M\to
\frac{M}{\tilde{\lambda}^2}. \ee  For the BTZ black hole, the ADM
mass is determined to be $M=\frac{r_+^2}{\ell^2}$.

Before we proceed, it is necessary to derive all thermodynamic
quantities. First of all, we mention that the Hawking temperature
can be determined from the metric by itself as
\begin{equation} \label{temp}
 T^{z}_H(r_+) =
 \frac{1}{4\pi}\Big[\sqrt{-g^{tt}g^{rr}}~ \left|g_{tt}'(r)\right|\Big]_{r=r_+}
     = \frac{r^z_+}{2\pi\ell^{z+1}},
\end{equation}
irrespective of knowing other conserved quantities. According to the
Euclidean action approach in~\cite{GTT}, one has to use the original
Bergshoeff-Hohm-Townsend action together with Gibbons-Hawking and
counter terms to obtain the $z=3$ regularized action \be
\label{regact} I_{reg}=I_{BHT}+I_{GH}+I_{ct}. \ee Here $I_{BHT}$ is
the Euclidean version of Bergshoeff-Hohm-Townsend action
as~\cite{bht} \be I_{BHT}=-\frac{1}{16 \pi G_3}\int_{{\cal M}} d^3x
\sqrt{-g} \Bigg[ {\cal R}-2\lambda +f^{\mu\nu}{\cal
G}_{\mu\nu}+\frac{m^2}{4}\Big(f_{\mu\nu}f^{\mu\nu}-f^2\Big)\Bigg],\ee
where $f^{\mu\nu}$ is an auxiliary field and ${\cal
G}_{\mu\nu}={\cal R}_{\mu\nu}-{\cal R}g_{\mu\nu}/2$.  The remaining
terms are boundary terms. The  Gibbons-Hawking term $I_{GH}$ \be
I_{GH}=-\frac{1}{16 \pi G_3}\int_{\partial {\cal M}} d^2x \sqrt{-h}
\Bigg[ -2K -\hat{f}^{ij}K_{ij}+\hat{f}K\Bigg]\ee  is  required to
have a well-defined  variational problem for the graviton. Here
$h_{ij}$ is an induced metric on the boundary, $K_{ij}$ is an
extrinsic curvature tensor and
$\hat{f}^{ij}=f^{ij}+2f^{r(i}N^{j)}+f^{rr}N^iN^j$ with the shift
$N^j$. Finally,  the counter-term $I_{ct}$ is necessary to
regularize the divergence on the boundary at infinity. It is given
by \be I_{ct}=\frac{1}{32 \pi G_3}\int_{\partial {\cal M}} d^2x
\sqrt{-h} \Bigg[ 15+\frac{\hat{f}}{2}-\frac{\hat{f}^2}{16}\Bigg].\ee
In this approach, the Euclidean time ($\tau=it$) is periodic as
$0\le \tau \le \beta$ and $0 \le \phi \le 2\pi \ell$.
 These are computed  to be
\be I_{BHT}=\frac{\beta r_+^2}{G_3 \ell^4}\Big(-r^2+r_+^2\Big),~~
I_{GH}=\frac{\beta r_+^2}{G_3 \ell^4}\Big(2r^2-r_+^2\Big),~~
I_{ct}=\frac{\beta r_+^2}{G_3
\ell^4}\Big(-r^2+\frac{3}{4}r_+^2\Big). \ee Adding these and thus,
divergences of $r\to \infty$  cancel out so that  the  Euclidean
action (\ref{regact}) is finite as \be \label{react}
I^{off}_{reg}=\frac{3\beta}{4G_3}\frac{r_+^4}{\ell^4}, \ee which
corresponds to the off-shell free energy.  In order to obtain the
on-shell free energy, we replace $\beta$ by $\beta^{z=3}_H(r_+)$ the
inverse of Hawking temperature (\ref{temp}) in (\ref{react}) which
leads to \be I^{on}_{reg}(r_+)=\frac{3}{4G_3}(2\pi
\ell)^{4/3}\frac{1}{(\beta^{z=3}_H)^{1/3}}. \ee It could be
rewritten as \be \label{onfree}
I^{on}_{reg}(r_+)=-\beta^{z=3}_H(r_+) F^{on}_{Lif}(r_+),~~
F^{on}_{Lif}(r_+)={\cal M}(r_+)-T^{z=3}_H(r_+)S^{z=3}_{BH}(r_+). \ee
By requiring that the action possess an extremum,  the mass ${\cal
M}$ is determined by \be {\cal M}(r_+)=-\frac{\partial }{\partial
\beta^{z=3}_H} I^{on}_{reg}. \ee On the other hand, replacing
$T^{z=3}_H$ by $T$ in the second of (\ref{onfree}) leads to the
off-shell free energy \be F^{off}_{Lif}(r_+,T)={\cal M}-T
S^{z=3}_{BH}.\ee   Here $T$ is an independent control parameter for
the study of phase transition. The on-shell (extremum) condition of
$ \frac{d}{dr_+} F^{off}_{Lif}=0$ (equivalently, requiring the
first-law of $d{\cal M}/dr_+=TdS^{z=3}_{BH}/dr_+$) determines $T$ to
be  the Hawking temperature \be T \to T^{z=3}_{H} \ee which shows
how the off-shell free energy reduces to the on-shell free energy:
\be F^{off}_{Lif}(r_+,T)|_{T\to T_H^{z=3}}\to F^{on}_{Lif}(r_+). \ee

For the $z=3$ Lifshitz black hole, its mass, heat capacity
($C=\frac{d{\cal M}}{dT_H^{z=3}}$), Bekenstein-Hawking entropy, and
on-shell (Helmholtz) free energy are given by \be\label{lbh}{\cal
M}=\frac{r_+^4}{4G_3\ell^4},~C=\frac{4\pi
r_+}{3G_3},~S^{z=3}_{BH}=\frac{2\pi
r_+}{G_3},~F^{on}_{Lif}=-\frac{3r_+^4}{4G_3\ell^4}.\ee Here we check
that the first-law of thermodynamics is satisfied as \be d{\cal
M}=T_H^{z=3}dS^{z=3}_{BH}. \ee On the other hand, the Lifshitz
soliton has its thermodynamic quantities~\cite{GTT} \be \label{lsol}
 T^{TL}_H=0,~{\cal M}_{TL}=-\frac{3}{4G_3},~S_{BH}=0,~F_{TL}=-\frac{3}{4G_3}.\ee because of the absence of horizon.
It corresponds to the manifold with the dynamical exponent $z=1/3$.

We mention the $z=1$ BTZ black hole case. For this case,
thermodynamic quantities are expressed in terms of a different
radial coordinate $\rho_+$
 \be \label{btz} M_{BTZ}=\frac{\rho_+^2}{8G_3\ell^2},~C=\frac{\pi \rho_+}{2G_3},~S^{z=1}_{BH}=
 \frac{\pi \rho_+}{2G_3},~F_{BTZ}= M_{BTZ}-T^{z=1}_H S^{z=1}_{BH}=-\frac{\rho_+^2}{8G_3\ell^2}.\ee
  Also the first-law of thermodynamics is
given by \be dM_{BTZ}=T_H^{z=1}dS^{z=1}_{BH}. \ee The thermal
AdS$_3$ has its thermodynamic quantities \be \label{tads}
M_{TAdS}=-\frac{1}{8G_3},~F_{TAdS}=-\frac{1}{8G_3}.\ee because of
the absence of horizon.
 At this stage, we discuss global structures of
 Lifshitz and BTZ black holes. Their Penrose diagram are similar as
 $\boxtimes$
 except that  the BTZ (Lifshitz) black holes are regular
 (singular) at $r=0$ (top and bottom), while at $r=\infty$ (two sides) the BTZ
 (Lifshitz) black hole spacetimes have asymptotically AdS (Lifshitz).

Now, we introduce  five parameters to study  phase transition in
black
hole physics~\cite{York}: \ba &&r_+ \to {\rm order~ parameter}, \nonumber \\
&&T_H(r_+) \to {\rm order~ parameter~(onshell~temperature)}, \nonumber \\
&&T \to {\rm control~ parameter~(offshell~temperature)}, \nonumber \\
&&F^{on}(r_+) \to {\rm increasing~(decreasing)~black~hole~via~equilibrium~process}, \nonumber \\
&&F^{off}(r_+,T) \to{\rm
increasing~(decreasing)~black~hole~via~nonequilibrium~process},
\nonumber  \ea where off-shell (on-shell) mean equilibrium
(non-equilibrium) configurations. In general, the equilibrium
process implies a reversible process, while the non-equilibrium
process implies a irreversible process.  The off-shell free energy
means a generalized free energy which is similar to a
temperature-dependent scalar potential $V(\varphi,T)$ for a simple
model of thermal phase transition.

Generally, increasing black hole ($\longrightarrow$) is induced  by
absorbing radiations in the heat reservior, while decreasing black
hole ($\longleftarrow$) is done by Hawking radiations as evaporation
process.

In studying the phase transition, two important quantities are the
heat capacity $C$ which shows thermal stability (instability) for
$C>0(C<0)$ and free energy $F^{on}$ which indicates the global
stability (instability) for $F^{on}<0(F^{on}>0)$.  For the case of
positive heat capacity ($C>0$), one relevant quantity is just  the
free energy ``$F^{on}$". Here we would like to mention that all
black holes under consideration are  thermally stable  because all
their heat capacities are positive.

{\it Hence, the free energy plays a key role in studying phase
transition between two gravitational configurations if they have the
same asymptote. If they have different asymptotics, comparing  free
energies does not make sense.}

\subsection{Transition between thermal AdS$_3$ and BTZ black hole}
\begin{figure}[t!]
   \centering
   \includegraphics{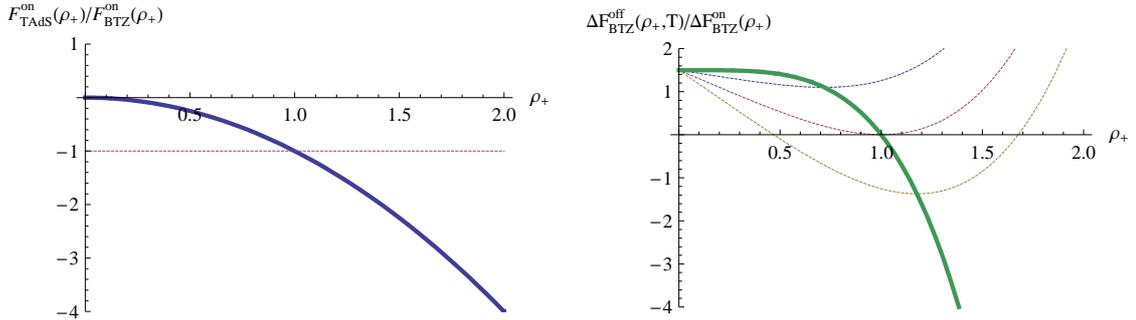}
\caption{Left panel: free energy of thermal AdS space [horizontal
line] and BTZ black hole [bold curve] with $G_3=1/8$ and $\ell=1$.
At $\rho_+=1$, $F^{on}_{TAdS}=F^{on}_{BTZ}=-1$. For
$\rho_+<1(\rho_+>1)$, the ground state is thermal AdS space (BTZ
black hole). Right: The bold curve denotes the difference in
on-shell free energy $\Delta F ^{on}_{BTZ}(\rho_+)$, while three
solid curves show the off-shell free energy $\Delta
F^{off}_{BTZ}(\rho_+,T)$ for three different temperatures: from top
to bottom, $T=0.059(<T_c),~T_c=0.159,~ 0.259(>T_c)$.}
\end{figure}
In order to discuss the phase transition, we first compare two free
energies of thermal AdS$_3$  (TAdS) and BTZ black hole.  One finds
from the left panel of Figure 1 that TAdS  is favored at small black
hole (low temperature), while BTZ black hole is favored at large
black hole (high temperature). This observation suggests  a phase
transition between TAdS  space and BTZ black hole. The
Horowitz-Myers conjecture for the AdS soliton~\cite{HM} implies that
the soliton with a negative energy could be taken as the thermal
background (ground state) in any dimensions.  We note that the
three-dimensional AdS soliton is just the thermal AdS$_3$ space
(TAdS)~\cite{SSW}.  Then,  we can calculate the difference of
 free energy with respect to TAdS as
 \be
\Delta
F^{on}_{BTZ}(\rho_+)=F^{on}_{BTZ}-F_{TAdS}=\frac{1}{8G_3}\Big[1-\frac{\rho_+^2}{\ell^2}\Big].
\ee  Also we introduce  the difference in  off-shell free energies
\be \Delta
F^{off}_{BTZ}(\rho_+,T)=F^{off}_{BTZ}(\rho_+,T)-F_{TAdS}=\frac{1}{8G_3}\Big[\frac{\rho_+^2}{\ell^2}+1\Big]-T
\cdot \frac{\pi \rho_+}{2G_3}.\ee   At the minimum point of
$\rho_+=\rho_m$ defined by the condition of
$dF^{off}_{BTZ}(\rho_+,T)/d\rho_+=0$, we have the relation of
$\Delta F^{on}_{BTZ}(\rho_m)=\Delta F^{off}_{BTZ}(\rho_m,T)$. This
is depicted in the right panel of Fig. 1.  At $T=T_c$, the
transition from TAdS to BTZ black hole is possible to occur. For
$T<T_c$, the TAdS dominates because of  $\Delta
F^{off}(\rho_m,T)>0$, while for $T>T_c$, the BTZ black hole
dominates because of $\Delta F^{off}(\rho_m,T)<0$.  This indicates a
change of dominance at the critical temperature $T=T_c=\frac{1}{2\pi
\ell}=0.159$~\cite{myungbtz}.

\begin{figure}[t!]
   \centering
   \includegraphics{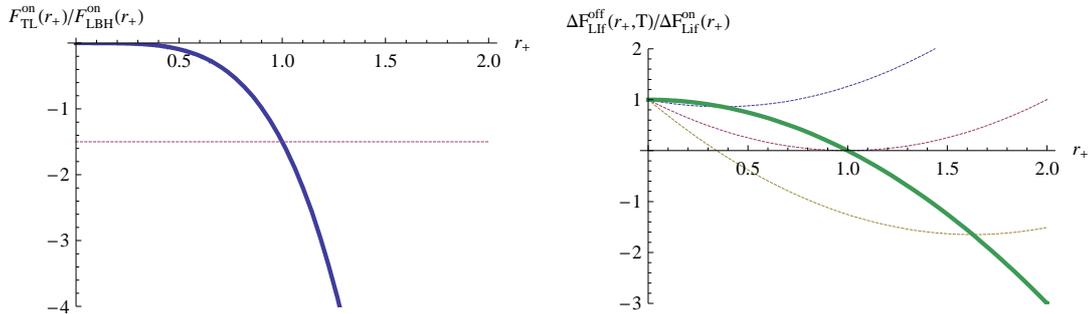}
\caption{Left: free energy of  Lifshitz soliton  [horizontal line]
and Lifshitz black hole [bold curve] with $G_3=1/2$ and $\ell=1$. At
$r_+=1$, $F^{on}_{TL}=F^{on}_{LBH}=-1$. For $r_+<1(r_+>1)$, the
ground state is  Lifshitz soliton (Lifshitz black hole). Right: The
bold curve denotes the difference in on-shell free energy $\Delta F
^{on}_{Lif}(r_+)$, while three solid curves show the off-shell free
energy $\Delta F^{off}_{Lif}(r_+,T)$ for three different
temperatures: from  top to bottom, $T=0.059(<T_c),~T_c=0.159,~
0.259(>T_c)$.}
\end{figure}
However, we wish to mention that this transition is not considered
as  a truly Hawking-Page transition because an unstable small black
hole with negative heat capacity was  missed in this
transition~\cite{Myungtads}. See Appendix for the Hawking-page
transition.

\subsection{Lifshitz soliton and Lifshitz black hole}
We observe the similarity between AdS systems of (\ref{btz}) and
(\ref{tads}) and Lifshitz systems of  (\ref{lbh}) and (\ref{lsol}).
Hence, we would like to compare two free energies of Lifshitz
soliton with
 $z=1/3$~\cite{GTT} and Lifshitz black hole with
$z=3$.  One finds from the left  of Figure 2 that Lifshitz  soliton
is favored at small black hole (low temperature), while Lifshitz
black hole is favored at large black hole (high temperature). This
observation may allow a phase transition between  Lifshitz soliton
and Lifshitz black hole.  The Lifshitz soliton with a negative
energy could be taken as the ground state, suggesting that the
Lifshitz soliton may be considered as  the thermal Lifshitz (TL).

 Now,
we can calculate the difference of
 free energy with respect to the Lifshitz soliton background  as
 \be
\Delta
F^{on}_{Lif}(r_+)=F^{on}_{LBH}-F_{TL}=\frac{3}{4G_3}\Big[1-\frac{r_+^4}{\ell^4}\Big].
\ee  We  define the difference in  off-shell free energies \be
\Delta
F^{off}_{Lif}(r_+,T)=F^{off}_{Lif}(r_+,T)-F_{TL}=\frac{1}{4G_3}\Big[\frac{r_+^4}{\ell^4}+3\Big]-T
\cdot \frac{2\pi r_+}{G_3} .\ee   At the minimum point of $r_+=r_m$,
we have $\Delta F^{on}_{Lif}(r_+)=\Delta F^{off}_{LiF}(r_+,T)$. This
is depicted in right of Figure 2.

 At the critical temperature $T=T_c$, a transition from the
 Lifshitz soliton to  Lifshitz black hole is possible to occur. For
$T<T_c$, the  Lifshitz soliton dominates because of $\Delta
F^{off}_{Lif}(r_m,T)>0$, whereas for $T>T_c$, the Lifshitz black
hole dominates because of $\Delta F^{off}_{Lif}(r_m,T)<0$. This may
indicate a change of dominance at the critical temperature
$T=T_c=\frac{1}{2\pi \ell}=0.159$, as the transition  occurs from
thermal AdS$_3$ space to BTZ black hole.  However,  this transition
seems not to occur  because their asymptotes  are
different~\cite{GTT}. Explicitly, the Lifshitz soliton  takes the
asymptote of ``$ - r^2 dt^2 + dr^2/r^2 + r^{2z} d\phi^2$", while the
Lifshitz black hole takes its asymptote of ``$ - r^{2z} dt^2 +
dr^2/r^2 + r^2 d\phi^2$".  They are the same only for $z=1$ AdS$_3$
spacetimes, but they are different for $z=3$ Lifshitz spacetimes.
Hence, in order to define  the phase transition properly,   we have
to find a new ground state of thermal Lifshitz with the Lifshitz
asymptote ($ - r^{2z} dt^2 + dr^2/r^2 + r^2 d\phi^2$).

Similarly, one may consider that a phase transition between Lifshitz
and BTZ black holes occur  because  their free energies in Figure 3
imply  that Lifshitz black hole is favored at small black hole (low
temperature), while BTZ black hole is favored at large black hole
(high temperature).
\begin{figure}[t!]
   \centering
   \includegraphics{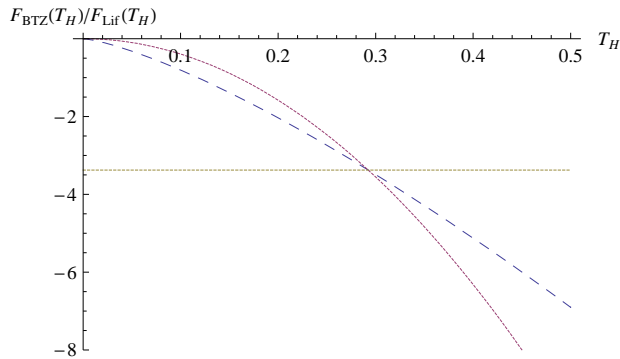}
\caption{Two free energies of Lifshitz [dashed] and BTZ [solid]
black holes with temperature matching  $T^{BTZ}_H=T^{LBH}_H$.  At
$T_H=T_c=0.29$, one finds that
$F^{on}_{Lif}(T_H)=F^{on}_{BTZ}(T_H)=-3.375$. It is suggested that
the ground state is Lifshitz black hole (BTZ black hole) for
$T_H<T_c~(T_H>T_c$). }
\end{figure}
However, there is no transition between Lifshitz and BTZ black holes
in three dimension because of their  different asymptotes of
Lifshitz and AdS$_3$ spaces.

\section{Lifshitz brane and black brane}
In order to confirm  ``no phase transition between Lifshitz black
hole and other configuration in three dimensions", we wish to study
possibility of  a phase transition between Lifshitz brane and black
brane in four dimensions.   For this purpose, we introduce the
effective action in four-dimensional spacetimes~\cite{tay}
\begin{equation} \label{lba}
S=\frac{1}{16\pi G_{4}}\int
d^{4}x\sqrt{-g}[R-2\Lambda-\frac{1}{2}\partial_{\mu}\phi\partial^{\mu}\phi
-\frac{1}{4}e^{\lambda\phi}F_{\mu\nu}F^{\mu\nu}],
\end{equation}
where $\Lambda$ is the cosmological constant and two fields are a
massless scalar and a Maxwell field. It admits the Lifshitz (black)
brane with dynamical exponent  $z=2$ as solution to  equations of
motion~\cite{pang2}
\begin{eqnarray}
\label{2eq2} &
&ds^{2}_{LB}=L^{2}\Big[-r^{2z}f(r)dt^{2}+\frac{dr^{2}}{r^{2}f(r)}+r^{2}\sum\limits^{2}_{i=1}
dx^{2}_{i}\Big],\nonumber\\
& &f(r)=1-\frac{r_{+}^{z+2}}{r^{z+2}},~~~e^{\lambda\phi}=\frac{1}{r^{4}},~~~\lambda^{2}=\frac{4}{z-1},\nonumber\\
&
&F_{rt}=qr^{z+1},~~~\Lambda=-\frac{(z+1)(z+2)}{2L^{2}},\nonumber\\
& &q^{2}=2L^{2}(z-1)(z+2),
\end{eqnarray}
where the event horizon is located at $r=r_{+}$. This line element
is invariant under the anisotropic scaling of $t\to
\tilde{\lambda}^zt,x_i \to \tilde{\lambda} x_i, r\to
r/\tilde{\lambda}$, and $r_+\to r_+/\tilde{\lambda}$. It is
important to note from the last relation that the charge $q$ is not
an independent charge hair because it is determined by the curvature
radius $L$ of Lifshitz black brane and its dynamic exponent $z$, in
compared  to the Reissner-Nordstr\"om-AdS black
hole~\cite{Chamblin:1999tk}. A similar case was found in the charged
MTZ black hole~\cite{cmtz,mp}

 On the
other hand,  the black brane (BB) solution  is obtained for $z=1$ as
\begin{eqnarray}
&
&ds^{2}_{BB}=L^{2}\Big[-r^{2}f(r)dt^{2}+\frac{dr^{2}}{r^{2}f(r)}+r^{2}\sum\limits^{2}_{i=1}
dx^{2}_{i}\Big],\nonumber\\
& &f(r)=1-\frac{r_{+}^{3}}{r^{3}},~~~\phi=\phi_{0}={\rm
const},\nonumber\\
& &F_{rt}=0,~~~\Lambda=-\frac{3}{L^{2}},
\end{eqnarray}
where the scalar and Maxwell field play no role. Hence,  the scalar
field $\phi$ and the Maxwell field $F_{\mu\nu}$ play the essential
role in modifying asymptotic geometry from AdS$_4$ spacetimes to
Lifshitz spacetimes.

The temperature and entropy are determined by
\begin{equation}
\label{2eq3}
T^z_H=\frac{1}{\beta^z_H}=\Big[\frac{z+2}{4\pi}\Big]r^{z}_{+},~S_{\rm
BH}=\frac{L^{2}V_{2}}{4G_{4}}r^{2}_{+},
\end{equation}
where $V_{2}$ denotes the volume of  two-dimensional spatial
directions. In order to use the Euclidean action approach, we need
 the Euclidean version $I_{E}$ of (\ref{lba}), \be
I_E=-\frac{1}{16\pi G_{4}}\int_{{\cal M}}
d^{4}x\sqrt{-g}[R-2\Lambda-\frac{1}{2}\partial_{\mu}\phi\partial^{\mu}\phi
-\frac{1}{4}e^{\lambda\phi}F_{\mu\nu}F^{\mu\nu}]\ee and the
Gibbons-Hawking term (extrinsic boundary term)~\cite{tay} \be
I_{GH}=-\frac{1}{8\pi G} \int d^3x \sqrt{h} K. \ee Also the counter
term (intrinsic boundary term) is given by \be
I_{ct}=\frac{z+1}{8\pi G} \int d^3 x \sqrt{h}. \ee Then, the
on-shell action of $I^{on}=I_E+I_{GH}+I_{ct}$ with $\beta=\beta^z_H$
leads to \be \label{eact} I^{on}=-\frac{V_2 L^2r_+^{z+2}}{16\pi G}
\beta^z_H=\beta^z_H F^{on}_{LB}. \ee
 Mass and heat capacity are obtained as \be M^z=\frac{2L^2V_2 r_+^{z+2}}{16\pi
G_4},~C^z=\frac{dM^z}{dT^z_H}=\frac{2V_2r_+^2}{4zG_4}=\frac{2S_{BH}}{z}>0.
\ee Because of positiveness of heat capacity for $z=1$ and 2, it is
natural to define an  on-shell free energy.

According to the Euclidean action approach, the on-shell free energy
can be read off from (\ref{eact}) as  \be
F_{LB}^{on}(r_+)=-\frac{L^2V_2 r_+^{z+2}}{16\pi G_4}, \ee which is
consistent with  the Gibbs free energy
\begin{equation} \label{Gibbs}
\tilde{F}^{on}_{LB}=M^z-\Phi Q-T^z_H S_{BH}=\frac{L^2V_2}{16\pi G_4}
\Big[2r_+^{z+2}+(z-1)r_+^{z+2}-(z+2)r_+^{z+2}\Big]=-\frac{L^2V_2
r_+^{z+2}}{16\pi G_4},\end{equation} with charge $Q=qV_2/32\pi G_4$
and potential $\Phi=-q r_+^{z+2}/(z+2)$. Here $q^2=2L^2(z-1)(z+2)$
was used to obtain $\tilde{F}^{on}_{LB}$.  We note  that  the
first-law of thermodynamics is satisfied to be  \be
dM^{z=2}=T^{z=2}_HdS_{BH}+\Phi dQ \to dM^{z=2}=T^{z=2}_HdS_{BH}\ee
because of $dQ=0$  for the $z=2$ Lifshitz brane.

In this case,   their off-shell free energies  are defined as  \ba
&&F^{off}_{LB}(r_+,T)=M^{z=2}-\Phi Q-T S_{BH}=3r_+^4-4\pi r_+^2 T,
\nonumber \\
&&F^{off}_{BB}(\rho_+,T)\mid_{\rho_+ \to
\frac{4}{3}r_+^2}=\Big[M^{z=1}-TS_{BH}\Big]_{\rho_+ \to
\frac{4}{3}r_+^2}=\Big(\frac{4}{3}\Big)^3\Big[2r_+^6-3\pi r_+^4
T\Big], \ea where  we used the normalization of $\frac{L^2V_2}{16\pi
G_4}=1$  for numerical computation and temperature matching of
$T^{z=2}_H=T^{z=1}_H$, implying that $\rho_+ \to (4/3)r_+^2$ in the
black brane sector. However,  the isotropic scaling ($z=1$) is
transformed to the anisotropic scaling ($z=2$)  when working with
the temperature matching.  As is shown in Figure 4, it is clear that
$dF^{off}_{LB}/dr_+=0 \nrightarrow T=T^{z=2}_H$, while
$dF^{off}_{BB}/dr_+=0 \rightarrow T=T^{z=1}_H$. In other words,
three minimum  points of off-shell Lifshitz free energy are not
crossed by on-shell free energy, while three minimum points of
off-shell black brane free energy are crossed by on-shell free
energy.

\begin{figure}[t!]
   \centering
   \includegraphics{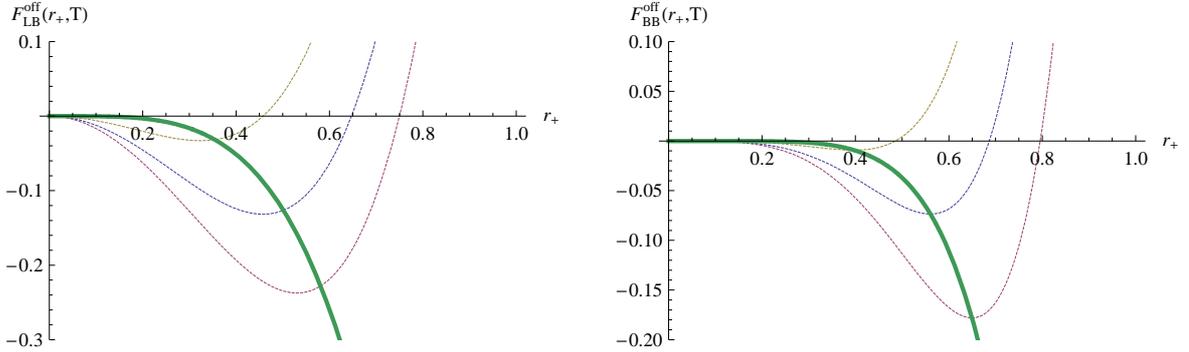}
\caption{Left: free energy of Lifshitz brane [on-shell: bold curve ;
off-shell: solid]. Three minimum points are not crossed by  on-shell
free energy.  This means that $F^{off}_{LB}$ is not suitable  for
describing the Lifshitz brane. Right: free energy of black brane
[on-shell: bold curve; off-shell: solid]. Here three minimum points
are crossed by on-shell free energy. This implies  that
$F^{off}_{BB}$ is  suitable  for describing for the black brane. }
\end{figure}

In order to have a better picture, on may use the Helmholtz free
energy $\bar{F}^{on}_{LB}(r_+)$ defined by \be \label{Helm}
\bar{F}^{on}_{LB}(r_+)=M^z-T^z_H S_{BH}=-\frac{zL^2V_2
r_+^{z+2}}{16\pi G_4} \ee because  $q^2$ is no longer an independent
charge hair. In this case, the first-law is satisfied as \be
dM^z=T^z_H dS_{BH}. \ee Thus, the corresponding off-shell free
energy \be \label{barf}
\bar{F}^{off}_{LB}(r_+,T)=M^z-TS_{BH}=2r_+^4-4\pi r_+^2 T \ee
 implies   that
 \be
 \frac{d}{dr_+}\bar{F}^{off}_{LB}(r_+,T)=0  \to T=\frac{dM^z}{dS_{BH}}    \to  T=T_H^z. \ee

We wish to mention   why the correct definition of the free energy
 is (\ref{Helm}) but  not (\ref{Gibbs}). Following Ref.~\cite{Tarrio}, the reason is just to have
a well defined variational problem at the boundary. As was  pointed
out after (\ref{2eq2}), the charge $q^2$ associated to the U(1)
gauge field is fixed by $L$ and $z$. This means that $q^2$  does not
correspond to the value of the gauge field at the boundary, but it
corresponds to  gradient of gauge field  which  must be kept fixed
when doing variations of the action to find the equations of motion.
Hence, choosing the Helmholtz free energy  amounts to changing
boundary conditions from the Dirichlet to the Neumann.  This could
be taken into account by  Legendre transformation which  leads to
the definition  (\ref{Helm}) for the free energy.

 On the other hand, it is well-known  that one uses the
Gibbs free energy when working in the grand canonical ensemble with
fixed  potential, while one uses the Helmholtz free energy working
in  the canonical ensemble with fixed charge. Hence there is a
problem in studying thermodynamics of  the Lifshitz brane,  which
states that fixing charge is not compatible with grand canonical
ensemble.

 As was suggested by Ref.~\cite{tay},
the anisotropic background of Lifshitz brane is favored at small
brane (low temperature), while the isotropic background of black
brane is favored at large brane (high temperature). It  is shown in
Figure 5.  This is mainly  because  $F^{on}_{LB}<F^{on}_{BB}$ for
$r_+<r_c=0.65$ and $F^{on}_{LB}>F^{on}_{BB}$ for $r_+>r_c=0.65$.
However, the suggested transition could not occur because their
asymptotes are different.
\begin{figure}[t!]
   \centering
   \includegraphics{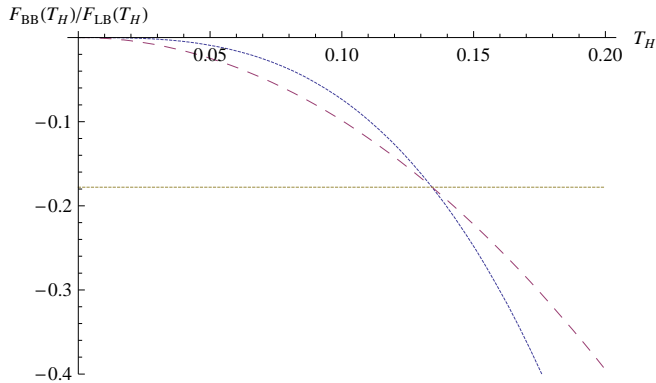}
\caption{Comparison of two free energies [Lifishtz brane (dashed)
and black brane (solid)] with temperature matching
$T^{z=2}_H=T^{z=1}_H$. At $T_H=T_c=0.13$, one finds that
$F^{on}_{LB}(T_H)=F^{on}_{BB}(T_H)=-0.18$. It is suggested that the
ground state is Lifshitz brane (black brane) for
$T_H<T_c~(T_H>T_c$). }
\end{figure}
\section{Discussions}

We have discussed  phase transitions between Lifshitz black holes
and other configurations by using free energy.  We summarize the
phase transitions: thermal AdS$_3$ $\to$ BTZ black hole, but
Lifshitz soliton $\nrightarrow$ Lifshitz black hole,  Lifshitz black
hole $\nrightarrow$ BTZ  black hole, and Lifshitz brane
$\nrightarrow$ black brane.

We remind the reader that  a necessary condition for a phase
transition to take place is  to have the same asymptote. In this
sense,  the above non-occurrences ($\nrightarrow$) are consistent
with the necessary condition for a phase transition in the
gravitational system. In other words,  the non-occurrence  is mainly
due to different asymptotes: asymptotically AdS and Lifshitz.

Consequently, if two configurations have the same asymptotes, the
free energy analysis is necessary to test whether does the phase
transition between two configurations occur. If not, the
conventional free energy analysis seems to be meaningless.
\newpage
\section*{Appendix: Hawking-page transition}

\begin{figure}[t!]
   \centering
   \includegraphics{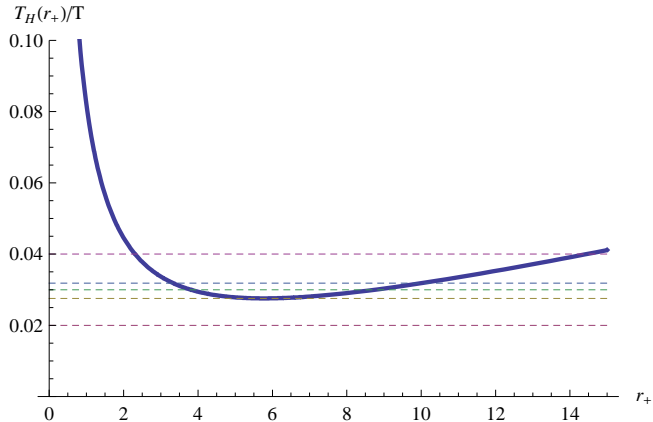}
\caption{Hawking temperature $T_H(r_+)$ (solid) and external five
temperatures $T=T_1(=0.02),~T_0(=0.027),~T_2(=0.03),~T_c(=0.031),$
and $T_3(=0.04)$. Here $T_{0}$ is the minimum temperature and $T_c$
is the critical temperature  for the Hawking-Page transition.}
\label{fig.10}
\end{figure}
In this appendix, we present the  Hawking-page transition as the
first-order phase transition in the Schwarzschild-AdS black hole
(SAdS)~\cite{HP}. The ADM mass, Hawking temperature, and the
Bekenstein-Hawking entropy are given by
\begin{equation}
\label{aas2}
M_{SAdS}(r_+)=\frac{1}{2}\Big(r_++\frac{r_+^3}{l^2}\Big),~~T_H(r_+)
= \frac{1}{4\pi} \Big( \frac{1}{r_+} + \frac{3r_+}{l^2}
\Big),~~S_{BH}=\pi r_+^2.
\end{equation}
As is shown in Figure 6, the shape of Hawking temperature is
$\smile$. The minimum temperature is $T_0=0.027$ for fixed $l=10$
and thus, it is impossible for a phase transition to take place for
$T<T_0$. The critical temperature is determined by the condition of
on-shell free energy.

The heat capacity and on-shell free energy are given
by~\cite{Myung:2008eb}
\begin{eqnarray}\label{aaSc}
C_{SAdS}(r_+)&=& 2\pi r_+^2 \Bigg(\frac{3r_+^4+l^2 r_+^2}{3r_+^4 - l^2 r_+^2}\Bigg), \\
\label{aaSf} F^{on}_{SAdS}(r_+)&=& M_{SAdS}-T_H S_{BH}=
\frac{r_{+}}{4}\Big(1 - \frac{r_{+}^2}{l^2}\Big),
\end{eqnarray}
where $C_{SAdS}$ blows up at $r_+=r_0=l/\sqrt{3}$ (heat capacity is
changed from $-\infty$ to $\infty$ at $r_+=r_0$) and
$F^{on}_{SAdS}=0$ for $r_+=r_c=l$. The thermal radiation is located
at $r_+=0$ in this black hole picture.
\begin{figure}[t!]
   \centering
   \includegraphics{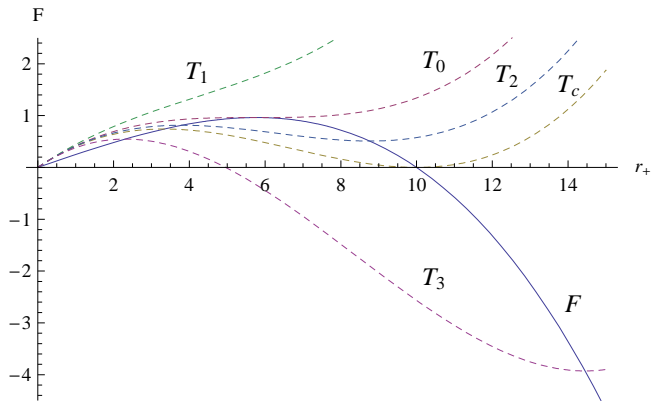}
\caption{Phase transition for the SAdS: the solid curve represents
the on-shell free energy $F^{on}_{SAdS}(r_+)$, while five dashed
curves denote the off-shell free energy $F_{SAdS}^{off}(r_+,T)$ with
five temperatures $T=T_1,~T_0,~T_2,~T_c,$ and $T_3$. }
\label{fig.10}
\end{figure}
 A SAdS  is globally stable only if  $C_{SAdS}>0$ and
$F^{on}_{SAdS}<0$.  We observe that the free energy is   maximum at
$r_+=r_0$ and zero at $r_+=r_c$ which determines the critical
temperature as shown in Figure 7. For $r_+>r_c$, one finds negative
free energy. The corresponding temperatures  are given by the
minimum $T_0=T_H(r_0)$ and the critical one $T_c=T_H(r_c)$,
respectively. Introducing the off-shell free energy as a function of
$r_+$ and $T$
\begin{equation}
F^{off}_{SAdS}(r_+,T)=M_{SAdS}(r_+)-T \cdot S_{BH}(r_+)
\end{equation}
we find the phase transition. Here $T$ plays a role of control
parameter for taking a phase transition.

Before we proceed, we point out an important relation \be
\frac{d}{dr_+}F^{off}_{SAdS}(r_+,T)=0 \to T=T_H \to
F^{off}_{SAdS}(r_+,T)|_{T\to T_H}=F^{on}_{SAdS}(r_+), \ee which
shows clearly how the off-shell free energy goes to the on-shell
free energy.  In other words, summing over all extremum  (saddle and
minimum points) in off-shell free energy $F^{off}_{SAdS}(r_+,T)$
provides the  on-shell free energy $F^{on}_{SAdS}(r_+)$.

 For $T=T_3>T_c$, the process of phase
transition is shown in Figure 7 explicitly.  In this case, one
starts with thermal radiation ($r_+=0$) in AdS space, a small black
hole (SBH$_-$)  appears  at $r_+=r_u$ [solution to
$F^{on}_{SAdS}(r_u)=F_{SAdS}^{off}(r_u,T_3)]$.  Here the SBH$_-$
denotes the unstable small black hole with $C_{SAdS}<0$ and
$F^{on}_{SAdS}>0$. Then, since the heat capacity changes from
$-\infty$ to $\infty$ at $r_+=r_0$, the large black hole (LBH$_+$)
finally comes out as a stable object at $r_+=r_s$ [solution to
$F^{SAdS}(r_s)=F^{SAdS}_{off}(r_s,T_3)]$. Here the LBH$_+$ denotes a
globally stable  black hole because of $C_{SAdS}>0$ and
$F^{on}_{SAdS}<0$.

 Actually, there is a change of the dominance at the critical
temperature $T=T_c$: from thermal radiation  to black hole~\cite{HP}
as seen from $F^{off}_{SAdS}(0,T_c)=F_{SAdS}^{off}(r_c,T_c)=0$. This
is called the  Hawking-Page  phase transition as a typical example
of the first-order transition in the gravitational system: thermal
gas $\to$ SBH$_-$ $\to$ LBH$_+$.  For two temperatures $T=T_1$ and $
T_2(<T_c)$, the free energy $F^{on}_{SAdS}(0)=0$ of thermal gas  is
the lowest one, while for the $T=T_3(>T_c)$ case, the lowest one is
the free energy $F^{on}_{SAdS}(r_s)<0$ for the large black hole.
Hence, for $T<T_c$, the ground state is thermal gas,  whereas for
$T>T_c$, the ground state is a large black hole. This dictates a
thermal phase transition which is controlled by the temperature $T$
in the gravitational system.

\section*{Acknowledgement}

We would like to thank Javier Tarrio for pointing out correctness of
free energy in the Lifshitz brane.  This work was supported by the
2012 Inje University research grant.

\end{document}